\documentclass[twocolumn,english,superscriptaddress, nofootinbib]{revtex4-2}
\usepackage[T1]{fontenc}
\usepackage[latin9]{inputenc}
\usepackage{geometry}
\usepackage{color}
\usepackage{units}
\usepackage{amsmath}
\usepackage{amssymb}
\usepackage{graphicx}

\makeatletter
\usepackage[colorlinks,citecolor=blue,
urlcolor=blue,hypertexnames=true,linkcolor=blue]
{hyperref} 

\usepackage{amssymb}
\usepackage{bbm}

\usepackage[colorlinks,citecolor=blue,
urlcolor=blue,hypertexnames=true,linkcolor=blue]
{hyperref} 
\usepackage[toc,page]{appendix}

\usepackage{etoolbox}
\appto\appendix{\addtocontents{toc}{\protect\setcounter{tocdepth}{0}}}

\appto\listoffigures{\addtocontents{lof}{\protect\setcounter{tocdepth}{1}}}
\appto\listoftables{\addtocontents{lot}{\protect\setcounter{tocdepth}{1}}}

\makeatother

\usepackage{babel}
\begin{document}
\title{{\Large{}Nonlinear quantum interferometric spectroscopy with entangled
photon pairs}}
\author{Shahaf Asban}
\email{sasban@uci.edu}

\affiliation{Department of Chemistry and Physics \& Astronomy, University of California,
Irvine, California 92697-2025, USA}
\author{Vladimir Y. Chernyak}
\email{chernyak@chem.wayne.edu}

\affiliation{Department of Chemistry, Wayne State University, 5101 Cass Ave, Detroit,
Michigan 48202, USA}
\affiliation{Department of Mathematics, Wayne State University, 656 W. Kirby, Detroit,
Michigan 48202, USA}
\author{Shaul Mukamel}
\email{smukamel@uci.edu}

\affiliation{Department of Chemistry and Physics \& Astronomy, University of California,
Irvine, California 92697-2025, USA}
\begin{abstract}
We develop closed expressions for a time-resolved photon counting
signal induced by an entangled photon pair in an interferometric spectroscopy
setup. Superoperator expressions in Liouville-space are derived that
can account for relaxation and dephasing induced by coupling to a
bath. Interferometric setups mix matter and light variables non-trivially,
which complicates their interpretation. We provide an intuitive modular
framework for this setup that simplifies its description. Based on
separation between the detection stage and the light-matter interaction
processes. We show that the pair entanglement time and the interferometric
time-variables control the observed physics time-scale. Only a few
processes contribute in the limiting case of small entanglement time
with respect to the sample response, and specific contributions can
be singled out.\textcolor{red}{{} }
\end{abstract}
\maketitle

\section{Introduction}

Interferometric setups introduce a promising new methodology for quantum
inference of matter information in quantum spectroscopy \citep{Asban_2021,Asban_2021_Sci,Dorfman_2021,Eshun_2021,Kushing_2020}.
Since quantum probes change their state upon interaction with external
systems, multiphoton coincidence detection schemes should reveal quantum
correlations induced by the sample \citep{Mukamel_2020,Asban_2019a}.
We consider the optical measurement setup shown in Fig. $\text{\ref{fig 0}}$
-- that includes linear and nonlinear elements that transform the
optical field before and after it is coupled to a sample. We focus
on time-resolved detection of two photons. We use one interferometric
setup to prepare the initial quantum state of the light, followed
by a second interferometer that manipulates the arrival times of matter
induced radiation to obtain matter pathway resolution. Interferometric
schemes such as the Mach-Zehnder \citep{rar90}, Hong-Ou-Mandel \citep{Hong_1987},
and Franson \citep{ray13} provide a useful toolbox to scan the change
in photon statistics by coupling to a sample from which matter information
may be inferred using multiphoton detection in coincidence \citep{kal08,ray13,Lavoie_2020}.
Quantum-enhancements of interferometric detection schemes are an experimental
reality in several fields \citep{Grice_1997}. The direct detection
of gravitational waves \citep{Caves_1981,Caves_1985,LIGO_2019}, unprecedented
phase estimation precision with high loss tolerance at lower photon
flux \citep{Hudelist_2014,Li_2014,Anderson_2017,Manceau_2017,Shaked_2018},
and in wide-field imaging \citep{Frascella_2019} are all contemporary
examples. Here, we introduce a new family of signals applied to the
inference of objects at the microscopic scale.

Liouville-space pathways break-down the density operator evolution
into the set of physical processes determined by time ordered excitation
and de-excitation processes (pathways) induced by the applied fields.
Sorting them out is important in order  to develop understanding the
underlying matter dynamics. This description become crucially important
when considering the effects of the environment that break time reversal
symmetry. Sorting these pathways allows to infer the role of each
process in a systematic manner. Liouville pathway resolution enable
to compare model-based theoretical predictions with experiment \citep{Mukamel_1995}. 

Our goal is to sort out the Liouville-space pathways, by scanning
interferometric delays in the preparation and detection stages. We
wish to identify what type of information regarding the dynamics of
the system and its coupling to a bath can be inferred from these measurements.
We consider two interferometers, at the preparation and detection
processes, as depicted in Fig. $\text{\ref{fig 0}}$ and further discussed
in Sec. $\text{\ref{sec:The-setup}}$. Two-port linear interferometers
induce transformations in the two-photon space \citep{Mota_2004,Mota_2016}.
They generate rotations in the basis of the electromagnetic field
\citep{Asban_2021}. These transformations offer a unique set of control
knobs used in both the preparation and detection stages. These control
parameters provide novel spectral windows \citep{Asban_2021_Sci}.
We show that the interferometric control variables enable to conveniently
scan the temporal dynamics. We derive compact superoperator expressions
for the time-resolved coincidence-signal of two photons expanded in
terms of Liouville-space pathways. For simplicity, we consider two
ideal detectors which is fast in comparison to all relaxation times
of the sample. This extends Glauber's celebrated theory of detection
\citep{Glauber_1963}. 

The general expressions are given in appendix B. For simplicity we
have derived an approximation for the limiting case of vanishing two-photon
temporal separation (the entanglement time $T_{e}$). The two photons
then arrive simultaneously relative to the observed dynamics timescale.
This corresponds to a vanishing moving time average of the response,
thus sensitive to the dynamics above $T_{e}$ which is in the femtosecond
regime. At this time scales we expect environment effects to be more
pronounced which requires the Liouville space approach. Moreover,
using the novel time variables, we are able to separate different
processes in the evolution of the sample in time domain, from which
different transport mechanisms can be studied in detail such as exciton-exciton
scattering \citep{Abramavicius_2009,Scholes_2017,Lee_2007}. 

\section{The setup \label{sec:The-setup}}

\begin{figure*}
\noindent \begin{centering}
\includegraphics[scale=0.47]{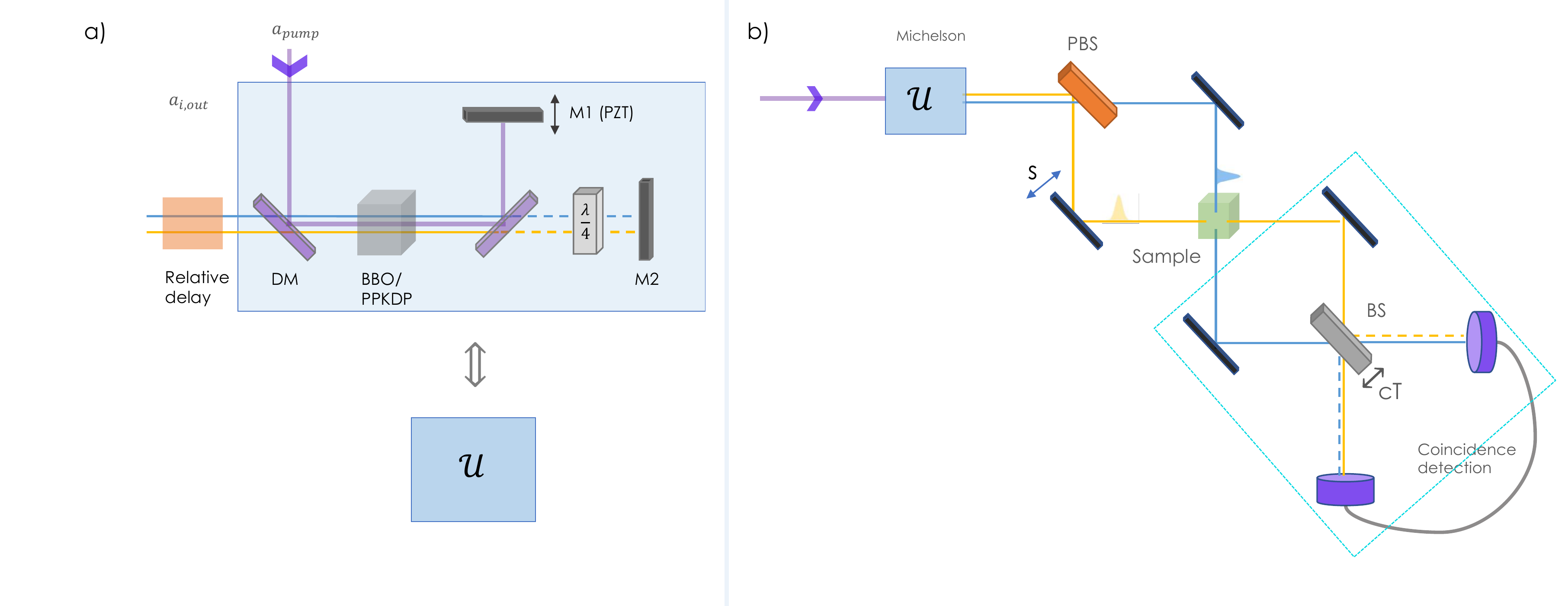}
\par\end{centering}
\caption{\textbf{The quantum interferometry setup}.(a) The Michelson interferometer
${\cal U}$ generates a pair of photons with variable exchange-phase
$\theta$, used as a control for the photon-pair degree of distinguishability
(see Eq. $\text{\ref{Theta symm. Amplitude}}$). The photons acquire
the relative delay $s$. (b) State preparation protocol in which two
photons are generated with a possible relative delay, then coupled
to thee sample and finally detected in coincidence in a Hong-Ou-Mandel
interferometer (boxed area). \label{fig 0}}

\end{figure*}

Our setup shown in Fig. $\text{\ref{fig 0}}$a contains two interferometers.
An entangled pair is prepared using a Michelson interferometer as
introduced in \citep{Grice_1997,Branning_1999,Branning_2000} and
depicted in Fig. $\text{\ref{fig 0}}$b. At the detection stage we
employ a Hong-Ou-Mandel interferometer. In this section we provide
the theoretical framework required for the inclusion of both interferometers.

\begin{figure*}
\includegraphics[scale=0.55]{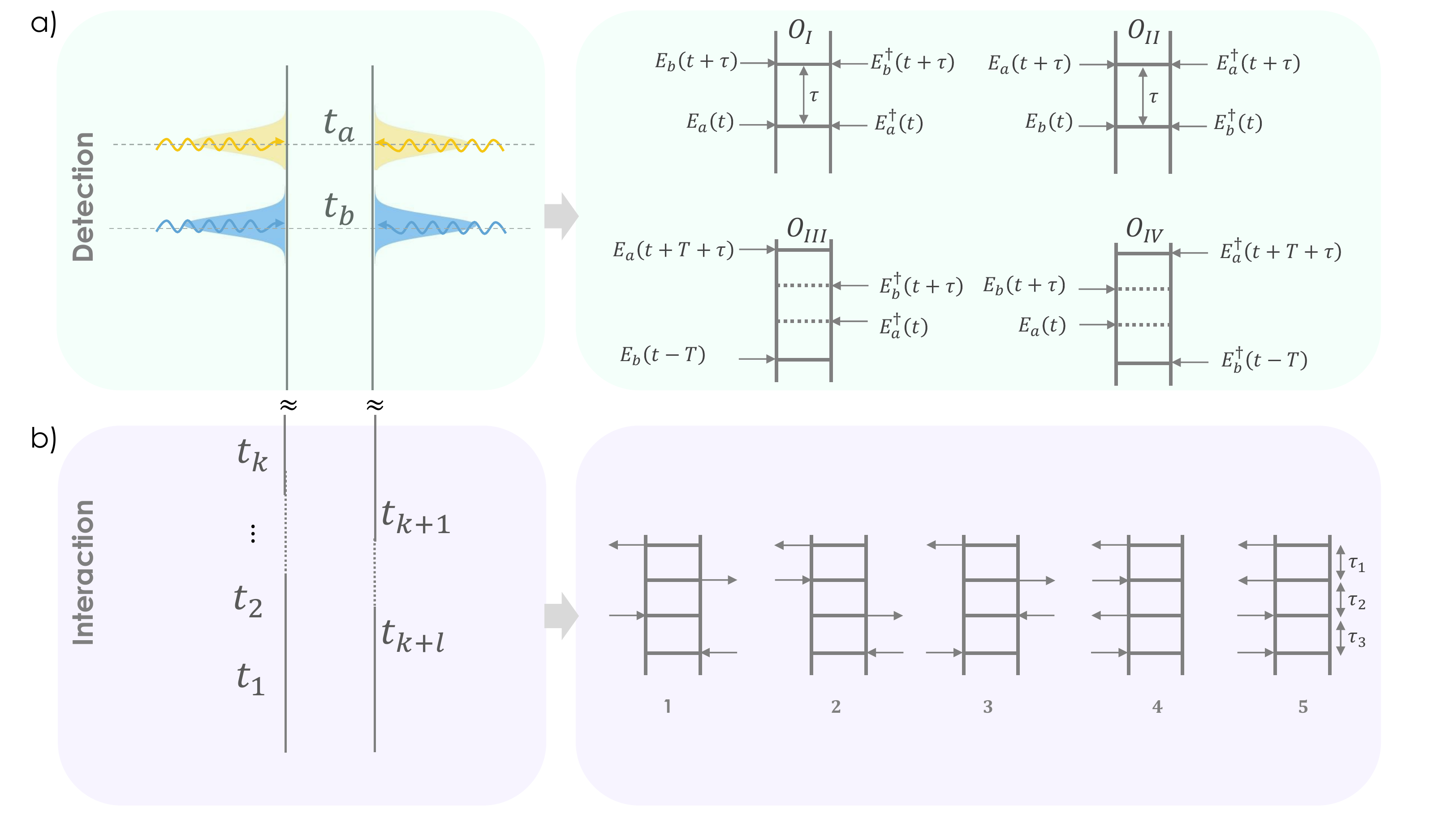}

\caption{\textbf{Block design for two-photon interferometric spectroscopy}.
Decoupling of the interaction and detection process when both are
well temporally defined. (a) The detection pathways corresponding
to HOM interferometer placed after the sample. Here $\tau>0$ is the
detection time difference of the two photons and $T>0$ is the HOM
delay. Two paths of direct propagation $O_{I-II}$, along with $O_{III-IV}$
in which there is a coherence between reflected and transmitted modes.
(b) Ladder diagrams that correspond to the light-matter interaction
corresponding to the matter correlation functions in Eq. $\text{\ref{eq:correlation functions}}$
(for a three-level model system). Only processes in which two-photons
interact with the sample and two are detected and sketched. \label{fig2}}
\end{figure*}

Each stage in the setup, introduces a superposition of fields which
we denote basis rotation or transformation. These platforms are presented
in a modular manner such that each stage is responsible for a well
defined property (e.g., introducing a delay). Quantum mechanically,
the stages are inseparable since more than one absorption, emission
and detection events occur in all possible time orderings. However,
we limit the discussion to a distant detection plane, such that the
propagation time of the emitted photons is much longer than the typical
time of the entire process. The detection and interaction are then
completely factorized temporally as depicted in Fig. $\text{\ref{fig2}}$ 

\subsection{Interferometric state preparation }

We assume a broadband ultrafast pump pulse which is known to imprint
identifying spectral information on each of the optical modes. The
modified Michelson interferometer in Fig. $\text{\ref{fig 0}}$a,
generates control over this feature by controlled systematization
of the wavefunction in a type-II phase matching setting. This results in engineered degree of distinguishability
as discussed below. 

The pump beam is reflected by the dichroic mirror (DM) then passes
the first time through the BBO crystal. The generated entangled pair
passes through the $\nicefrac{\lambda}{4}$ plate that switches the
polarization such that $\phi\left(\omega_{a},\omega_{b}\right)\rightarrow\phi\left(\omega_{b},\omega_{a}\right)$,
then exits the interferometer from the left. Otherwise, the pump photon
is reflected by the second DM and passes through the BBO crystal for
the second time with a controlled phase introduced by the PZT device
on which mirror M1 is positioned generating possibly an entangled
pair. Finally the pump beam is filtered out of the interferometer
by the first DM. 

The combined nonlinear interferometer transformations create a two
photon wavefunction of the form

\begin{equation}
\vert\Psi_{\theta}\rangle=\int d\omega_{a}d\omega_{b}\Phi_{\theta}\left(\omega_{a},\omega_{b}\right)a^{\dagger}\left(\omega_{a}\right)b^{\dagger}\left(\omega_{b}\right)\vert\text{vac}\rangle.\label{eq: wavefunction}
\end{equation}

\noindent The amplitude is given by

\noindent 
\begin{equation}
\Phi_{\theta}\left(\omega_{a},\omega_{b}\right)=\frac{1}{\sqrt{2}}\left[\phi\left(\omega_{a},\omega_{b}\right)+e^{i\theta}\phi\left(\omega_{b},\omega_{a}\right)\right].\label{Theta symm. Amplitude}
\end{equation}
The joint spectral amplitude (JSA) resulting from the direct-channel
$\phi\left(\omega_{a},\omega_{b}\right)$ (due to one pass in the
BBO with aligned polarizations) used in our calculations is given
by $\phi\left(\omega_{a},\omega_{b}\right)={\cal A}_{p}\left(\omega_{a}+\omega_{b}\right)\varphi\left(\omega_{a},\omega_{b}\right)$.
The Gaussian pump envelope ${\cal A}_{p}\left(\omega\right)=\exp\left[\nicefrac{\left(\omega-\omega_{p}\right)^{2}}{\sigma_{p}^{2}}\right]$
is centered around $\omega_{p}$ characterized with the bandwidth
$\sigma_{p}$ \citep{Law_2000}. The phase-matching factor $\varphi\left(\omega_{a},\omega_{b}\right)=\text{sinc}\left[\left(\omega_{a}-\bar{\omega}_{a}\right)T_{a}+\left(\omega_{b}-\bar{\omega}_{b}\right)T_{b}\right]$,
breaks the frequency exchange symmetry, i.e., $\varphi\left(\omega_{a},\omega_{b}\right)\neq\varphi\left(\omega_{b},\omega_{a}\right)$.
Here $\bar{\omega}_{a/b}$ is the central frequency of the signal
and idler beams, and $T_{a/b}=L\left(v_{a/b}^{-1}-v_{p}^{-1}\right)$,
where $L$ is the nonlinear crystal length and $v$ is the inverse
group velocity at the relevant central frequency $\left(\bar{\omega}_{a/b},\omega_{p}\right)$. In this (type-II) phase matching condition, the channels are flipped using the their opposite  and orthogonal polarization degree of freedom, e.g., $\vert \Psi_\theta\rangle=\vert HV\rangle + e^{i\theta}\vert VH\rangle$ where $V$ and $H$ correspond to horizontal and vertical polarizations respectively. For different phase matching condition which give rise to identiclly polarized biphotons, one would expect a different output, e.g., $\vert \Psi_\theta\rangle=\vert HH\rangle + e^{i\theta}\vert VV\rangle$.

Here, we have calculated the preparation step in the Sch\"odinger
picture, modifying the initial amplitude rather than the fields operators.
The detection stage is computed using the Heisenberg picture as explained
in Sec. $\text{\ref{sec:Detection-pathways}}$. This way, the dynamics
is calculated in the matter's reference-frame, as done in \citep{Asban_2021_Sci}.
Each interferometer introduces additional spectroscopic control parameters
that can be used to study the joint light-matter quantum state.

\section{Detection pathways \label{sec:Detection-pathways}}

The detection process involves a HOM interferometer as shown in the
boxed area of Fig. $\text{\ref{fig 0}}$b. We define the signal in
time domain. We consider a sample described by the Hamiltonian $H_{\mu}$,
that is coupled to field degrees of freedom $H_{\phi}$ by the dipolar
interaction $H_{\text{int}}=\boldsymbol{\mu}\left(\boldsymbol{r},t\right)\cdot\boldsymbol{E}\left(\boldsymbol{r},t\right)$.
Here $\boldsymbol{\mu}\left(\boldsymbol{r},t\right)=\boldsymbol{V}\left(\boldsymbol{r},t\right)+\boldsymbol{V}^{\dagger}\left(\boldsymbol{r},t\right)$
is the dipole operator, $\boldsymbol{V}=\sum_{i<j}\mu_{ij}\vert i\rangle\langle j\vert$
is a lowering transition operator acting in the molecular Hilbert
space with the corresponding matrix element $\mu_{ij}$, $\boldsymbol{E}\left(\boldsymbol{r},t\right)=\sum_{\sigma}\left[E_{\sigma}\left(\boldsymbol{r},t\right)+\text{H.c. }\right]$
is the electric field operator given by $E_{\sigma}\left(\boldsymbol{r},t\right)=\sum_{\boldsymbol{k}}\sqrt{\frac{2\pi k}{\Omega_{Q}}}\hat{\epsilon}_{\sigma}\left(\boldsymbol{k}\right)a_{\boldsymbol{k},\sigma}\left(t\right)e^{i\boldsymbol{k}\cdot\boldsymbol{r}}$,
where $\hat{\epsilon}_{\sigma}\left(\boldsymbol{k}\right)$ is the
$\sigma$-polarization vector, $\Omega_{Q}$ is the quantization volume
($c=1$), and $a_{\boldsymbol{k},\sigma}\left(a_{\boldsymbol{k},\sigma}^{\dagger}\right)$
are (bosonic) photon annihilation (creation) operators obeying $\left[a_{\boldsymbol{k},\sigma},a_{\boldsymbol{k}',\sigma'}^{\dagger}\right]=\delta_{\sigma,\sigma'}\delta_{\boldsymbol{k},\boldsymbol{k}'}$.
Hereafter we assume that the applied field is near resonance with
a molecular transition, such that the rotating wave approximation
may be applied by setting $H_{\text{int}}=EV^{\dagger}+E^{\dagger}V$.

The two-photon coincidence signal is defined by 

\begin{align}
\hat{{\cal O}}\left(t,\tau\right) & =E_{a',R}^{\dagger}\left(\boldsymbol{r}_{a},t\right)E_{b',R}^{\dagger}\left(\boldsymbol{r}_{b},t+\tau\right)\label{eq: total count observable}\\
 & \times E_{b',L}\left(\boldsymbol{r}_{b},t+\tau\right)E_{a',L}\left(\boldsymbol{r}_{a},t\right).\nonumber 
\end{align}

\noindent Here, $E_{R}$ and $E_{L}$ are electric field superoperators
that act from the right $E_{R}\rho\equiv\rho E$, and left $E_{L}\rho\equiv E\rho$
of the density operator. We further define their linear combinations
$A_{\pm}O=A_{L}O\pm A_{R}O$ (corresponding to Hilbert-space commutator
and anti commutator). These will be useful for the compact description
in the interaction picture below. The primes reflect the fact that
the detection and interaction planes are described in different basis
sets due to the HOM transformation, as explained below. 

\subsection{The HOM detection interferometer}

In our setup in Fig. $\text{\ref{fig 0}}$, two beams are combined
on a beam splitter (BS) which is mounted on a movable stage. This
enables to scan variable propagation times of the reflected pathways
of the beam with respect to the transmitted ones, introducing the
HOM time-delay $T$. This process is described by a linear transformation
since the photons at the detection plane are represented in a different
basis. The transformation (Jordan-Schwinger map) can be represented
as an $SU\left(2\right)$ rotation in the frequency-domain \citep{Yurke_1986,Jauche_1976,Mota_2004,Mota_2004_2,Mota_2016},
resulting in the input-output relation for the field operators. Writing
the field in vector notation given by $\boldsymbol{E}\left(\boldsymbol{r},\omega\right)=\left(E_{a}\left(\boldsymbol{r},\omega\right),E_{b}\left(\boldsymbol{r},\omega\right)\right)^{T}$,
the HOM rotation the detected field is given by $\boldsymbol{E}\vert_{\text{detection}}\left(\boldsymbol{r},\omega\right)=\hat{{\cal R}}_{T}\boldsymbol{E}\vert_{\text{interaction}}\left(\boldsymbol{r},\omega\right)$
where $\hat{{\cal R}}_{T}$ is given by

\begin{equation}
\hat{{\cal R}}_{T}=\left(\begin{array}{cc}
t & ire^{i\omega T}\\
ire^{-i\omega T} & t
\end{array}\right).\label{HOM transf.}
\end{equation}

\noindent Here $t$ and $r$ are the transmission and reflection coefficients
obeying $\left|t\right|^{2}+\left|r\right|^{2}=1$. For the 50:50
BS considered here, $t=r=\nicefrac{1}{\sqrt{2}}$. In the following,
we express all field operators in the matter interaction-domain basis
$\boldsymbol{E}\vert_{\text{interaction}}\equiv\boldsymbol{E}\left(\boldsymbol{r},\omega\right)$,
which requires the inverse rotation of the observable in Eq. $\text{\ref{eq: total count observable}}$
\citep{Asban_2021}. 

\subsection{The observable}

Glauber's $G^{\left(2\right)}$$\left(\tau\right)$ coincidence signal
is formally given by the expectation value of this observable, evolved
using the total density matrix of the field and matter in the interaction
picture

\noindent 
\begin{align}
{\cal C}\left(t,\tau\right) & =\left\langle {\cal T}\hat{{\cal O}}^{\prime}\left(t,\tau\right)e^{-\frac{i}{\hbar}\underset{-\infty}{\overset{t^{*}}{\int}}dsH{}_{\text{int},-}\left(s\right)}\right\rangle ,\label{Signal def.}
\end{align}

\noindent here ${\cal T}$ is the time ordering superoperator that
maintains the bookkeeping of the interaction events, e.g., ${\cal T}A\left(t_{1}\right)B\left(t_{2}\right)=\theta\left(t_{1}-t_{2}\right)A\left(t_{1}\right)B\left(t_{2}\right)+\theta\left(t_{2}-t_{1}\right)B\left(t_{2}\right)A\left(t_{1}\right)$,
the Heaviside step-function is defined by $\theta\left(t\right)=1,$
$\forall t\geq0$ and $\theta\left(t\right)=0$, $\forall t<0$. Note
that the interaction Hamiltonian superoperator is represented by the
field modes prior to the transformation. HOMI introduces the time
delay $T$ as an additional control parameter to the observable superoperator
$O\left(t,\tau\right)\rightarrow O\left(t,\tau,T\right)$. For the
HOM detection setup, we should transform the observable in Eq. $\text{\ref{eq: total count observable}}$
according to the HOM transformation (Eq. $\text{\ref{HOM transf.}}$),
resulting in 16 detection pathways. Only the four -- in which one
photon of each mode are detected -- contribute to our signal (see
also \citep{Hong_1987}), reducing Eq. $\text{\ref{eq: total count observable}}$
to 

\noindent 
\begin{align}
\hat{{\cal O}}^{\prime}\left(t,\tau,T\right) & =O_{I}+O_{II}+O_{III}+O_{IV}\label{eq: Observable pathways}
\end{align}

\noindent where the detection pathways are given by 

\begin{subequations}

\noindent {\small{}
\begin{align}
O_{I} & =E_{a,R}^{\dagger}\left(\boldsymbol{r}_{a},t\right)E_{b,R}^{\dagger}\left(\boldsymbol{r}_{b},t+\tau\right)E_{b,L}\left(\boldsymbol{r}_{b},t+\tau\right)E_{a,L}\left(\boldsymbol{r}_{a},t\right)\label{eq: detection pathways}\\
O_{II} & =E_{b,R}^{\dagger}\left(\boldsymbol{r}_{a},t\right)E_{a,R}^{\dagger}\left(\boldsymbol{r}_{b},t+\tau\right)E_{a,L}\left(\boldsymbol{r}_{b},t+\tau\right)E_{b,L}\left(\boldsymbol{r}_{a},t\right)\\
O_{III} & =-E_{a,R}^{\dagger}\left(\boldsymbol{r}_{a},t\right)E_{b,R}^{\dagger}\left(\boldsymbol{r}_{b},t+\tau\right)\\
 & \times E_{a,L}\left(\boldsymbol{r}_{b},t+T+\tau\right)E_{b,L}\left(\boldsymbol{r}_{a},t-T\right)\nonumber \\
O_{IV} & =-E_{b,R}^{\dagger}\left(\boldsymbol{r}_{a},t-T\right)E_{a,R}^{\dagger}\left(\boldsymbol{r}_{b},t+T+\tau\right)\\
 & \times E_{b,L}\left(\boldsymbol{r}_{b},t+\tau\right)E_{a,L}\left(\boldsymbol{r}_{a},t\right).\nonumber 
\end{align}
}{\small\par}

\end{subequations}

Note that since Eq. $\text{\ref{eq: Observable pathways}}$ is given
in the basis of the interaction domain (different from Eq. $\text{\ref{eq: total count observable}}$),
none of the quantities are primed in the definition of $\hat{{\cal O}}$
as well as the field operators $E_{m,X}$ ($m\in\left\{ a,b\right\} $;
$X\in\left\{ L,R\right\} $). We have explicitly included the HOM
delay $T$ variable to the coincidence observable, which is expressed
in $O_{III-IV}$ and a $\left(-\right)$ sign {[}Eq. $\text{\ref{eq: Observable pathways}}${]}.
All four combinations depicted in Fig. $\text{\ref{fig2}}$a (top-right)
contribute to the interferometric coincidence signal. When the BS
is removed, the ordinary coincidence detection setup can also be recovered
by only keeping the $O_{I}$ contributions. 

\section{The interaction pathways}

\begin{figure}
\noindent \begin{centering}
\includegraphics[scale=0.45]{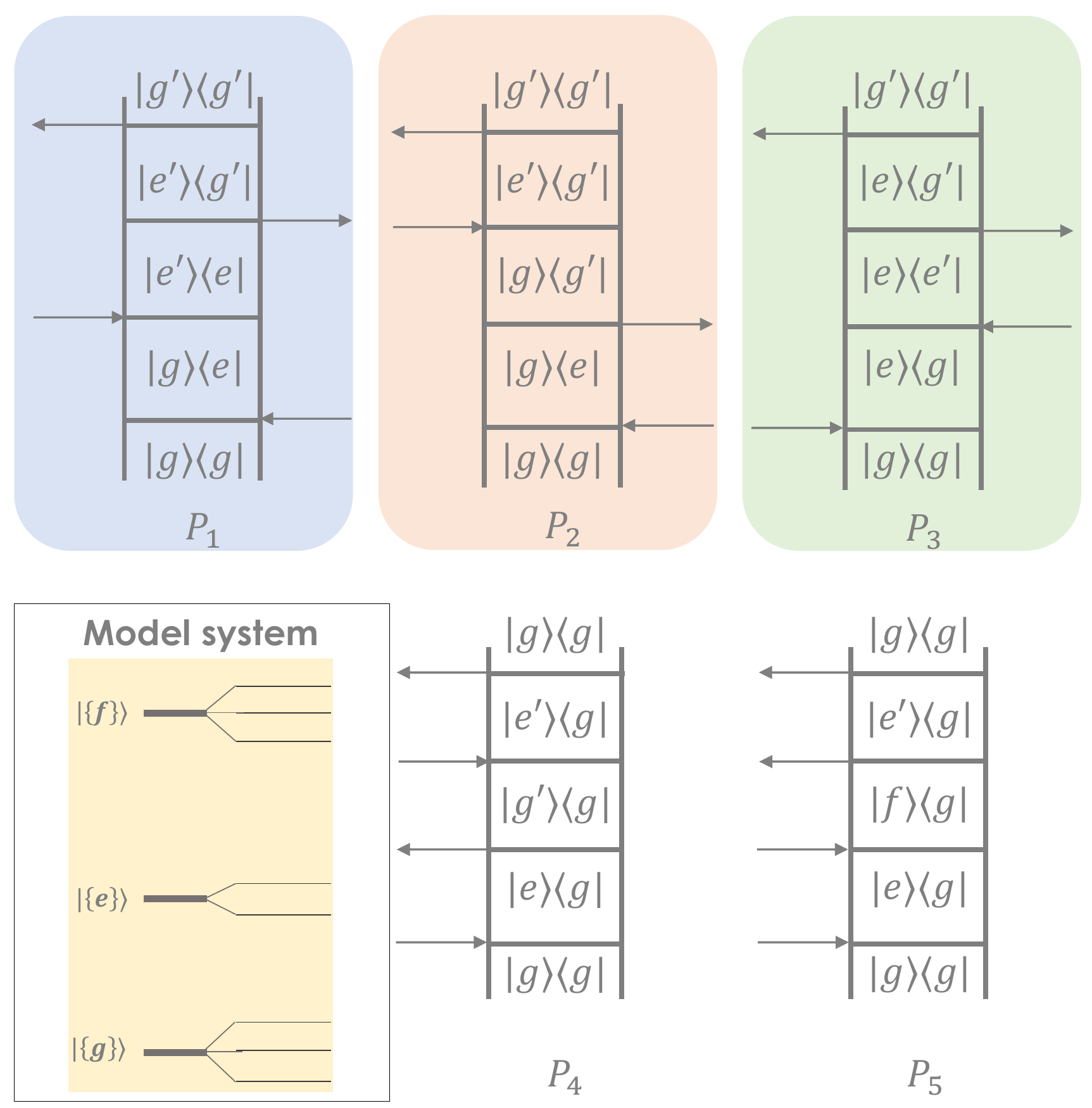}
\par\end{centering}
\caption{\textbf{Interaction pathways for three-level exciton model}. The five
interaction pathways with the respective exciton dynamics for the
selected three-level exciton model system. Each energy manifold $\vert\left\{ \boldsymbol{g}\right\} \rangle,\vert\left\{ \boldsymbol{e}\right\} \rangle,\vert\left\{ \boldsymbol{f}\right\} \rangle$
is composed of several states. \label{fig3}}

\end{figure}

We expand the signal in Eq. $\text{\ref{Signal def.}}$ pertubatively
to $4^{\text{th}}$ order in $H_{\text{int}}$ such that each photon
interacts twice with the sample. Generally, 4 interactions generate
16 left-right Liouville pathways. In addition, each arrow may point
inward/outward resulting in a total of 256 possible pathways. As depicted
in Figs. $\text{\ref{fig2}}$a and b, this number is significantly
reduced mainly due to the coincidence detection and the initial Fock
state. Note that Fig. $\text{\ref{fig2}}$b contains half of the contributions
and their complex conjugates should be added. It is possible to further
reduce the number of contributions, by the following considerations.
Near resonance, the rotating wave approximation (RWA) can be invoked,
resulting in the simplified interaction Hamiltonian $H_{\text{int}}=EV^{\dagger}+E^{\dagger}V$.
We further consider the three level model systems depicted in Fig.
$\text{\ref{fig3}}$, initially in the ground state $\rho_{0}=\vert g\rangle\langle g\vert$,
so that the first interaction can only be excitation (no de-excitation).
This eliminates contributions in which an emission event occurs after
a single photon is detected. The expectation value $\left\langle \hat{{\cal O}}\right\rangle $
is and expectation value and thus real (note that the diagrams in
Fig. $\text{\ref{fig3}}$ are symmetric with respect to exchange of
L-R and taking the complex conjugate). By convention, we only include
pathways in which the last interaction is taken from the left side
with an outgoing arrow (generated a detected photon). The contributions
in which the last interaction is from the right are related to these
by conjugation and interchanging L-R. The full signal is finally given
by $2\mathfrak{Re}\left\langle \hat{{\cal O}}p_{i}\right\rangle $
where $p_{i}$ denotes all pathways terminated at the left. The top
interaction from the left must point outwards, otherwise this mode
is not occupied, hence not detected. Photon number conservation implies equal number of inward/outward
arrows. Only diagrams in which two photons interact with the sample
and two photons are detected contribute to the signal. Since two-photon
population is detected, diagrams in which there is a single arrow
in one of the sides are eliminated. 

The five surviving pathways are depicted in Fig. $\text{\ref{fig3}}$,
all contain two field modes from each side of the density operator.
Their complex conjugates should be added as well. These processes
are labeled $P_{i}$ where $i\in\left[1,5\right]$ in Fig. $\text{\ref{fig3}}$,
correspond to the superoperator correlation functions denoted ${\cal F}_{i}$
given by 

\begin{subequations}

{\small{}
\begin{align}
{\cal F}_{1}\left(\tau_{1},\tau_{2},\tau_{3}\right) & =\left\langle V_{L}{\cal G}\left(\tau_{1}\right)V_{R}^{\dagger}{\cal G}\left(\tau_{2}\right)V_{L}^{\dagger}{\cal G}\left(\tau_{3}\right)V_{R}\right\rangle \label{eq:correlation functions}\\
{\cal F}_{2}\left(\tau_{1},\tau_{2},\tau_{3}\right) & =\left\langle V_{L}{\cal G}\left(\tau_{1}\right)V_{L}^{\dagger}{\cal G}\left(\tau_{2}\right)V_{R}^{\dagger}{\cal G}\left(\tau_{3}\right)V_{R}\right\rangle \\
{\cal F}_{3}\left(\tau_{1},\tau_{2},\tau_{3}\right) & =\left\langle V_{L}{\cal G}\left(\tau_{1}\right)V_{R}^{\dagger}{\cal G}\left(\tau_{2}\right)V_{R}{\cal G}\left(\tau_{3}\right)V_{L}^{\dagger}\right\rangle \\
{\cal F}_{4}\left(\tau_{1},\tau_{2},\tau_{3}\right) & =\left\langle V_{L}{\cal G}\left(\tau_{1}\right)V_{L}^{\dagger}{\cal G}\left(\tau_{2}\right)V_{L}{\cal G}\left(\tau_{3}\right)V_{L}^{\dagger}\right\rangle \\
{\cal F}_{5}\left(\tau_{1},\tau_{2},\tau_{3}\right) & =\left\langle V_{L}{\cal G}\left(\tau_{1}\right)V_{L}{\cal G}\left(\tau_{2}\right)V_{L}^{\dagger}{\cal G}\left(\tau_{3}\right)V_{L}^{\dagger}\right\rangle ,
\end{align}
}{\small\par}

\end{subequations}

\noindent (plus their complex conjugates). Here $\left\langle \cdots\right\rangle \equiv\text{tr}\left\{ \cdots\rho_{\mu}\left(-\infty\right)\right\} $
where $\rho_{\mu}\left(-\infty\right)$ is the initial state of the
matter, and Liouville-space Green's function is given by ${\cal G}\left(t\right)=-\frac{i}{\hbar}\theta\left(t\right)e^{-i{\cal L}_{\mu}t-\eta t}$. 

The matter correlation functions  ${\cal F}_{j} (\tau_{1}, \tau_{2}, \tau_{3})$,
with $j = 1, \ldots, 5$, presented in Eq. $\text{\ref{eq:correlation functions}}$
and illustrated in Fig. $\text{\ref{fig3}}$, provide a useful microscopic
insight into the capabilities of the entangled-photon spectroscopy
to retrieve detailed information on ultrafast photoinduced dynamics
of various chemical systems, inter alia molecular aggregates, whose
dynamics is determined by the electronic interactions induced Frenkel
exciton scattering and exciton-phonon interactions. Since the actual
measured correlated signals are represented by convolutions of ${\cal F}_{j}$
with the doorway and window functions, the latter containing relevant
information on the entangled photon sources, as well as interferometric
supplement, and having nothing to do with the dynamics of the system
under study, one should in first place understand what kind of a matter
dynamical information is contained in the aforementioned four-point
correlators. 

Since all matter dynamical information, available via coherent four-wave
mixing spectroscopy is fully contained in the third-order nonlinear
response function 

\begin{eqnarray} 
\label{F-chi-3} {\cal F}_{\rm coh} = \left\langle \hat{V}_{+} {\cal G} (\tau_{1}) \hat{V}_{-} {\cal G} (\tau_{2}) \hat{V}_{-} {\cal G} (\tau_{3}) \hat{V}_{-}\right\rangle, 
\end{eqnarray}  

\noindent with $\hat{V} = V + V^{\dagger}$, the differences in the
spectroscopic information, provided by entangled-photon versus four-wave
mixing spectroscopies, originates just from the different Liouville-space
structure of the four-point matter correlators ${\cal F}$. Although
the latter is apparently very different for ${\cal F}$, presented
in Eq. $\text{\ref{eq:correlation functions}}$ versus Eq. \ref{F-chi-3},
the issue allows for a clear, simple, and conceptually explicit analysis,
as follows.

Indeed, as shown, using Liouville space Green functions techniques
\citep{CHERNYAK_1995a,Meier_1997b} and further by means of the Nonlinear
Exciton Equations (NEE) \citep{Chernyak_1998}, in the case of weak
to moderate exciton-phonon coupling, when the dynamics is dominated
by excitonic effects, and effects of polaron formation/self-trapping
are not substantial, there are three phenomena that contribute to
optical response: (i) exciton-exciton scattering, described in terms
of the exciton-exciton scattering matrix $\bar{\Gamma} (\omega)$,
(ii) exciton-photon coupling mediated transport, described by the
exciton transport correlation function $G^{(2)} (\omega)$, and combined
effects, expressed in terms of a convolutions of the two above; of
course, the expressions for the response contain the one-exciton Green
function $G (\omega)$ that contains of the exciton-phonon coupling
induced exciton dephasing.

The Green function approach can be extended to analyze the Liouville
space correlators ${\cal F}_{j}$ in Eq.~(8); this analysis will be
addressed in detail in a separate publication, here we just present
its main outcome, to relate it to the Feynman diagrams (Fig. $\text{\ref{fig3}}$).
Despite a very different Liouville-space structure of the correlators
in the entangled photon versus four-wave mixing case, due to the specific
features of the Frenkel exciton model with moderate exciton-photon
coupling, the ingredients that enter the final expressions, namely
$\bar{\Gamma} (\omega)$, $G^{(2)} (\omega)$, and $G (\omega)$ stay
the same, just the final expressions get modified. Translating the
results, presented in Eq. $\text{\ref{eq:correlation functions}}$
from the Sum-over-States (SOS) to the exciton-scattering language,
we can combine the Feynman diagrams (4) and (5) in Fig. $\text{\ref{fig3}}$
to obtain type (i) effects, i.e., pure exciton-excitons scattering;
it is well known that combining these two diagrams we take care of
the so-called cancellation of the $N^{2}$ terms problem, which in
the exciton-scattering approach happens automatically. Combining the
diagrams (1), (2), and (3) we obtain a type (ii) contribution that
reflects exciton transport effects, since in diagrams (1) and (3)
the system is in the population/exciton-exciton coherence state during
the $\tau_{2}$ time segment. In diagram 2 the system is in the $|g\rangle \langle g'|$,
which means in the ground electronic state with a different phonon
structures, it plays a proper cancellation role for exciton transport,
in the way how the diagram (4) operates for exciton scattering. Note
that both diagrams (2) and (4) also describe slight modifiication/renormalization
of the excitons scattering matrix due to exciton-phonon coupling.

Very importantly type (iii) effects (combined exciton-exciton scattering
and exciton transport) in the four-wave mixing, on the SOS language
originate from the diagrams, when the system is in the population/exciton-exciton
coherence and two-exciton ground state coherence ($|f\rangle \langle g|$ or $|g\rangle \langle f|$)
during the time periods $\tau_{2}$ and $\tau_{3}$, respectively.
Since such diagrams never appear in the entangled-photon spectroscopy
case, type (iii) effects do not contribute to the signals in the latter
case.

Summarizing, unlike the coherent four-wave mixing spectroscopy, the
entangled-photon spectroscopy studies only effects of exciton-exciton
scattering and exciton transport, with no combined contributions,
thus provided a better separation of dynamical phenomena that contribute
to spectroscopic data.

\subsection{The signal}

In Fig. $\text{\ref{fig 0}}$a we consider the HOMI detection setup.
The preparation alters the primary source, resulting in the JSA given
in Eq. $\text{\ref{Theta symm. Amplitude}}$. The HOMI detection transforms
the $G^{\left(2\right)}\left(\tau\right)$ signal in Eq. $\text{\ref{eq: total count observable}}$
into Eq. $\text{\ref{eq: Observable pathways}}$.

The total signal involves summing over all the product combinations
of detection pathways in Fig. $\text{\ref{fig2} }$a with all the
interaction pathways depicted in Fig. $\text{\ref{fig2}}$b (see appendix
B for the expressions of all combinations for a general preparation
process). The signal contains many terms corresponding to all combinations
of Liouville pathways $i\in\left[1,5\right]$, with all detection
pathways defined in Eq. $\text{\ref{eq: Observable pathways}}$ $O_{\nu}$,
where $\nu=I,...,IV$. One way to think about it is that each process
$P_{i}$ is obtained by a coherent superposition of all HOM detection
pathways ${\cal S}_{\nu,i}$, and the signal is given by their superposition.
For an illustrative example of the derivation of an contribution,
see appendix B. The coincidence signal in Eq. $\text{\ref{eq: signal}}$
is finally given by 

\begin{align}
{\cal C}\left(\tau,T,s\right) & =2\mathfrak{Re}\left\{ \sum_{\nu,i}{\cal S}_{\nu,i}\right\} ,\label{eq: signal}
\end{align}

\noindent where ${\cal S}_{\nu,i}\equiv\text{tr}\left\{ O_{\nu}P_{i}\rho\left(-\infty\right)\right\} $
and the detection and interaction pathways are labeled in Fig. $\text{\ref{fig2}}$.
The detection pathways are given in Eq. $\text{\ref{eq: Observable pathways}}$
(see appendices for detailed expressions). Note that the density matrix
is given by a product of the matter and field respectively $\rho\left(-\infty\right)\equiv\rho_{\mu}\otimes\rho_{\varphi}$.
The field is traced with respect to $\rho_{\varphi}\equiv\vert\Psi_{\theta}\rangle\langle\Psi_{\theta}\vert$
using $\Phi_{\theta}\left(t_{1},t_{2}\right)=\frac{1}{\sqrt{2}}\left[\phi\left(t_{1},t_{2}\right)+e^{-i\theta}\phi\left(t_{2},t_{1}\right)\right]$
with respect to Eq. $\text{\ref{eq: wavefunction}}$ due to the Michelson
interferometer. 

\section*{The short entanglement-time limit}

We no invoke an approximation which greatly simplifies this signal.
Consider a symmetric joint spectral amplitude, obtained by either
using $\theta=0$ in Eq. $\text{\ref{Theta symm. Amplitude}}$, or
via a narrowband pump. In either case, the entanglement time $T_{e}$
represents the time window in which both photons arrive \citep{Branning_1999}.
We also consider the characteristic timescale for the matter dynamics
to be bound from above by $\tau_{\text{R}}$. We focus on the regime
$\tau_{R}\gg T_{e}$ such that both photons arrive simultaneously.
The relative delay $s$ introduced in Fig. $\text{\ref{fig 0}}$a
between the pair, now sets the time interval in which all interactions
occur. In this limit, the amplitude is approximated by a narrow distribution 

\begin{equation}
\Phi\left(t_{1},t_{2}\right)\rightarrow\delta\left(t_{1}-t_{2}-s\right).\label{eq:nar_band}
\end{equation}

\noindent Consequently, processes with vanishing time intervals between
interactions do not contribute to this order $\left(\tau_{1},\tau_{2},\tau_{3}\right)\neq0$.
Note that $\Phi\left(t_{1},t_{2}\right)$ is symmetric to $t_{1}\leftrightarrow t_{2}$
exchange under this approximation which is consistent with $\theta=0$.
Here the Michelson interferometer is used for rectification of the
exchange phase $\theta$ when an ultrafast pump is used. Alternatively,
a narrowband pump can be used in which the exchange phase correction
is no longer essential. Since $\left(T,\tau,s\right)$ are measured
on a finite grid, we define the discrete time delta distribution $\delta_{t_{1},t_{2}}$
which attains the value $1$ when $t_{1}=t_{2}$ within our setup.
Plugging Eq. $\text{\ref{eq:nar_band}}$, in the signal $\left(\text{Eq. }\text{\ref{eq: signal_sup}},\text{appendix}\right)$
for $\tau\geq-T$ and $s>0$ we obtain 

\onecolumngrid

\begin{align}
{\cal C}\left(\tau>-T,T,s\right)= & 2\text{\ensuremath{\mathfrak{Re}}\ensuremath{\left\{  -\theta\left(\tau+T\right){\cal F}_{1}\left(T,\tau+T-s,2s-T\right)\vert_{E}\right.}}\nonumber \\
 & -\theta\left(\tau+T\right){\cal F}_{2}\left(2T+\tau-s,s-T-\tau,\tau+s\right)\vert_{E}\nonumber \\
 & -\theta\left(\tau+T\right)\left[{\cal F}_{3}\left(T,\tau+s,T-2s\right)+{\cal F}_{3}\left(T+\tau,s-2T-\tau,T+\tau\right)\right]\vert_{E}\nonumber \\
 & +2\delta_{\tau,s}\int d\tau_{3}{\cal F}_{5}\left(\left|\tau\right|,\tau_{3},\left|\tau\right|\right)\vert_{D}\left.-2\delta_{2T+\tau,s}\int d\tau_{3}{\cal F}_{5}\left(\left|\tau\right|,\tau_{3},2T+\tau\right)\vert_{E}\right\} \label{eq:main apprx}
\end{align}

\noindent where the coincidence contribution ${\cal C}_{4}$ does
not appear in this limit. We have introduced the notation $\vert_{E/D}$
corresponding to the direct $\left(O_{I,II}\right)$ and exchange
$\left(O_{III,IV}\right)$ paths of the HOM interferometer. From this
we see that only process $i=5$ in Fig. $\text{\ref{fig3}}$ contributes
to both the direct paths $O_{I/II}$, the rest are limited to the
exchanges $O_{III/IV}$. In this limit we already appreciate the degree
of control offered by the interferometric setups, offering novel temporal
inference tool-box. For example, for $s=0$, only exchange path processes
may contribute since $s$ set up the scale in which all the interactions
with the sample occur. HOM exchange paths are not restricted by this
due to the ambiguity in the arrival times. We thus single out ${\cal F}_{3}$
contribution, ${\cal C}\left(\tau>-T,T,s=0\right)\propto-\text{\ensuremath{\theta\left(\tau+T\right)}}\mathfrak{Re}\left\{ {\cal F}_{3}\left(T,\tau,T\right)\right\} $.
Also, for $\tau=0$ we obtain

\begin{subequations}

\begin{align}
{\cal C}\left(\tau=0,s<T<2s,s\right) & \propto\theta\left(T\right)\text{\ensuremath{\mathfrak{Re}}\ensuremath{\left\{  {\cal F}_{1}\left(T,T-s,2s-T\right)\right\} } },\label{eq:Limiting-Signal}\\
{\cal C}\left(\tau=0,\frac{s}{2}<T<s,s\right) & \propto\theta\left(T\right)\text{\ensuremath{\mathfrak{Re}}\ensuremath{\left\{  {\cal F}_{2}\left(2T-s,s-T,s\right)\right\} } },\\
{\cal C}\left(\tau=0,2s<T,s\right) & \propto\theta\left(T\right)\text{\ensuremath{\mathfrak{Re}}\ensuremath{\left\{  {\cal F}_{3}\left(T,s,T-2s\right)\right\} } }.
\end{align}

\end{subequations}

\twocolumngrid

\noindent Eqs. $\text{\ref{eq:Limiting-Signal}-c}$ demonstrate that
by following the multidimensional data, it is possible to isolate
certain contributions in the time domain. 

\section{Discussion}

The setup above presents several types of control variable over the
signal. These can be categorized in three groups: (a) classical pump,
(b) preparation and (c) detection parameters. The pump related parameters
includes the central frequency of the pump and its spectral width
$\left(\omega_{p},\sigma_{p}\right)$. The preparation setup is rich
with parameters including the central frequencies of the daughter
photons and their respective time $\left(\bar{\omega}_{a},\bar{\omega}_{b},T_{a},T_{b}\right)$,
using the phase matching conditions and dispersion properties of the
nonlinear crystal at these frequencies (see Sec. $\text{\ref{sec:The-setup}}$).
These parameters were not scanned here and offer a rich playground
for future studies. The detection parameters include the number of
detected photons (here two) and the HOM Delay $T$. 

Interferometric spectroscopy with quantum light has several merits.
Due to the specified number of interacting and detected photons, certain
pathways which contribute to classical signals, are eliminated. This
feature has strictly quantum origin since we are using Fock states.
Due to the application and detection of fixed number of photons, the
signal records only processes that lie within the two-photon subspace.
This greatly reduce the number of Liouville-space pathways. One can
define an entropic measure from the Liouville pathways probability
functional which depends on the preparation and detection details.
Ultimately, it is possible to identify the activated pathways by shaping
this probability with the available control parameters. This represents
the quantum information gain obtained by the protocol. One way to
see that is by defining a pathway related entropy ${\cal S}_{0}=-\sum_{i=1}^{L}P_{i}\log P_{i}$
where $L$ is the number of Liouville pathways with a given probe
and $P_{i}$ is the probability of the $i^{\text{th}}$ pathway. Then
compare it to the entropy of the interferometric setup ${\cal S}=-\sum_{i=1}^{L}Q_{i}\log Q_{i}$
where $Q_{i}$ is the overall probability of the $i^{\text{th}}$
pathway in the manipulated scheme. Ultimately, one can quantify the
quantum inference due to the detection process alone using an identical
probe and different detection by calculating the Kullback-Liebler
divergence ${\cal D}\left[P||Q\right]=\sum_{i=1}^{L}p_{i}\log\frac{P_{i}}{Q_{i}}$
\citep{Kullback_1951}. 

Each of the delays in this setup affects the signal differently, which
allows to control the pathways in the time domain. For example, by
taking $\Phi\left(t_{1},t_{2}\right)\rightarrow\delta\left(t_{1}-t_{2}-s\right),$
the entire process duration is set by $s$ for the direct detection
pathways. This stems from the fact that one photon is absorbed and
emitted while its entangled partner goes straight to the detector
in some pathways. One observes a superposition of processes in which
both photon interact with the sample with a conjugate process in which
both had not. Therefore, the two-photon coherence time sets the characteristic
interaction duration for the direct pathways. The exchange detection
pathways have more freedom due to the superposition in time domain
introduced by the HOMI. When the material system under study possesses
several characteristic time-scales $\tau_{R}$, they can be studied
separately by adjusting $s$.

Coincidence-detection further involves unique scaling relations between
the applied intensity $I_{p}$, the light-sample coupling and the
detected signal. This allows to avoid damaging the sample by using
weak quantum fields \citep{Asban_2021_Sci}.
\begin{acknowledgments}
The support of the National Science Foundation (NSF) Grant CHE-1953045
is gratefully acknowledged. The research was supported by the U.S.
Department of Energy (DOE), Office of Science, Basic Energy Sciences,
under Award Number DE-SC0022134. S.M. and V.C. were supported by the
DOE award.
\end{acknowledgments}

\bibliographystyle{unsrt}
\bibliography{TDLS}

\pagebreak{}

\onecolumngrid

\appendix

\part*{}

\section{Hilbert-space approach to photon counting }

In this section we propose an alternative derivation in which the
entire calculation is computed in Hilbert space. This representation
has the advantage of offering more compact expressions reflected in
less diagrams. However, when external degrees of freedom are included
to account for inaccessible processes as a result of possible coupling
to the environment, this is no longer possible. 

Our observable is expressed via interaction of an field mode with
the detector, changing its polarization. This can be described using
perturbative expansion of the interaction Hamiltonian with the detector
degrees of freedom such that each interaction event contributes a
single interaction to the wavefunction that describes the light, sample
and detector. An $N$ photon measurement operator corresponds to \\
\begin{equation}
{\cal M}^{\prime}={\cal T}\int\cdots\int dt_{1}\cdot\dots dt_{N}d\boldsymbol{r}_{1}\cdot\dots d\boldsymbol{r}_{N}H_{\text{int}}\left(t_{1}\right)\cdot\dots H_{\text{int}}\left(t_{N}\right)
\end{equation}

\noindent where each active pixel correspond to a single interaction
Hamiltonian $H_{\text{int}}\left(t\right)=\boldsymbol{E}\cdot\boldsymbol{V}$.
When the detector's dipole response is taken to be small and fast,
we approximate it as delta distribution in space-time and obtain the
known Glauber detection scheme. We consider an ordered measurement
scheme without loss of generality for the two-photon coincidence scheme.
We also assume that the detection plane is far from the sample such
that the time ordering operator does not mix the two photon detection
with the light-matter coupling. The wavefunction of the sample and
the detectors at positions $\left\{ \boldsymbol{r}_{i},t_{i}\right\} $
is separable and therefore takes the form

\[
\vert\Theta\left(t\right);\left\{ \boldsymbol{r}_{i},t_{i}\right\} \rangle_{\mu\varphi M}=\vert\Psi\left(t\right)\rangle_{\mu\varphi}\vert g,g\rangle_{M},
\]

\noindent here he subscripts $M,\varphi,\mu$ describes the detectors,
field and sample respectively. After the interaction with the detectors 

\begin{align}
\vert\Theta\left(t\right);\left\{ \boldsymbol{r}_{i},t_{i}\right\} \rangle_{\mu\varphi M} & \underset{\text{2 detectors}}{=}E\left(\boldsymbol{r}_{a},t_{a}\right)E\left(\boldsymbol{r}_{b},t_{b}\right)\vert\Psi\left(t\right)\rangle_{\mu\varphi}\vert e_{1},e_{2}\rangle_{M}\nonumber \\
 & ={\cal M}\left(\boldsymbol{r}_{a},\boldsymbol{r}_{b};t_{a},t_{b}\right)\vert\Psi\left(t\right)\rangle_{\mu\varphi}\vert e_{1},e_{2}\rangle_{M},
\end{align}

\noindent Developing the light-sample wavefunction pertubatively we
obtain 

\begin{align}
\vert\Psi\left(t\right)\rangle_{\mu\varphi} & =\sum_{k=0}\vert\Psi^{\left(k\right)}\left(t\right)\rangle_{\mu\varphi}\\
\vert\Psi^{\left(k\right)}\left(t\right)\rangle_{\mu\varphi} & =\left(-\frac{i}{\hbar}\right)^{k}{\cal T}_{+}\int\cdots\int dt_{1}\cdot\dots dt_{k}d\boldsymbol{r}_{1}\cdot\dots d\boldsymbol{r}_{k}H_{\text{int}}\left(t_{1}\right)\cdot\dots H_{\text{int}}\left(t_{k}\right)\vert\Psi\left(-\infty\right)\rangle_{\mu\varphi},
\end{align}

\noindent where ${\cal T}_{+}$ denotes the time ordering operator
forward in time ${\cal T}_{+}A\left(t_{1}\right)B\left(t_{2}\right)=\theta\left(t_{1}-t_{2}\right)A\left(t_{1}\right)B\left(t_{2}\right)+\theta\left(t_{2}-t_{1}\right)B\left(t_{2}\right)A\left(t_{1}\right)$.
Note that we introduced the $\left(+\right)$ subscript to the time
ordering since the Hermitian conjugate evolves formally backwards
in time using ${\cal T}_{-}$ to the left of the observable. We calculate
the probability of this detection setup by taking the modulus square
of this amplitude, resulting in 

\begin{align}
{\cal P}\left(\left\{ \boldsymbol{r}_{i},t_{i}\right\} \right) & =\langle\Theta\left(t\right);\left\{ \boldsymbol{r}_{i},t_{i}\right\} \vert\Theta\left(t\right);\left\{ \boldsymbol{r}_{i},t_{i}\right\} \rangle_{\mu\varphi M},\\
 & =\langle\Psi\left(t\right){\cal M}^{\dagger}\left(\left\{ \boldsymbol{r}_{i},t_{i}\right\} \right){\cal T}_{-}\vert{\cal T}_{+}{\cal M}\left(\left\{ \boldsymbol{r}_{i},t_{i}\right\} \right)\Psi\left(t\right)\rangle_{\mu\varphi}
\end{align}

explicitly 

\begin{equation}
{\cal P}\left(\boldsymbol{r}_{a},\boldsymbol{r}_{b};t_{a},t_{b}\right)=\sum_{k,l=0}^{\infty}\langle\Psi^{\left(l\right)}\left(t\right)E^{\dagger}\left(\boldsymbol{r}_{b},t_{b}\right)E^{\dagger}\left(\boldsymbol{r}_{a},t_{a}\right){\cal T}_{-}\vert{\cal T}_{+}E\left(\boldsymbol{r}_{a},t_{a}\right)E\left(\boldsymbol{r}_{b},t_{b}\right)\Psi^{\left(k\right)}\left(t\right)\rangle_{\mu\varphi}.
\end{equation}

This equation is exact in the light-matter interaction and pertubative
in the interaction with the detector. Note that until this stage,
all field-source contractions are permitted such that emission of
a photon can occur after the detection of another. From this point,
we assume that the detectors are placed far from the interaction area
such that one can assume the time ordering applies for the light-matter
interaction solely and the detection events are ordered by definition
$\left(t_{a}>t_{b}\right)$ 

\begin{equation}
{\cal P}\left(\boldsymbol{r}_{a},\boldsymbol{r}_{b};t_{a},t_{b}\right)=\sum_{k,l=0}\langle\Psi^{\left(l\right)}\left(t\right)\vert E^{\dagger}\left(\boldsymbol{r}_{b},t_{b}\right)E^{\dagger}\left(\boldsymbol{r}_{a},t_{a}\right)E\left(\boldsymbol{r}_{a},t_{a}\right)E\left(\boldsymbol{r}_{b},t_{b}\right)\vert\Psi^{\left(k\right)}\left(t\right)\rangle_{\mu\varphi}.\label{eq: Coin_Prob}
\end{equation}

\begin{figure}
\begin{centering}
\includegraphics[scale=0.6]{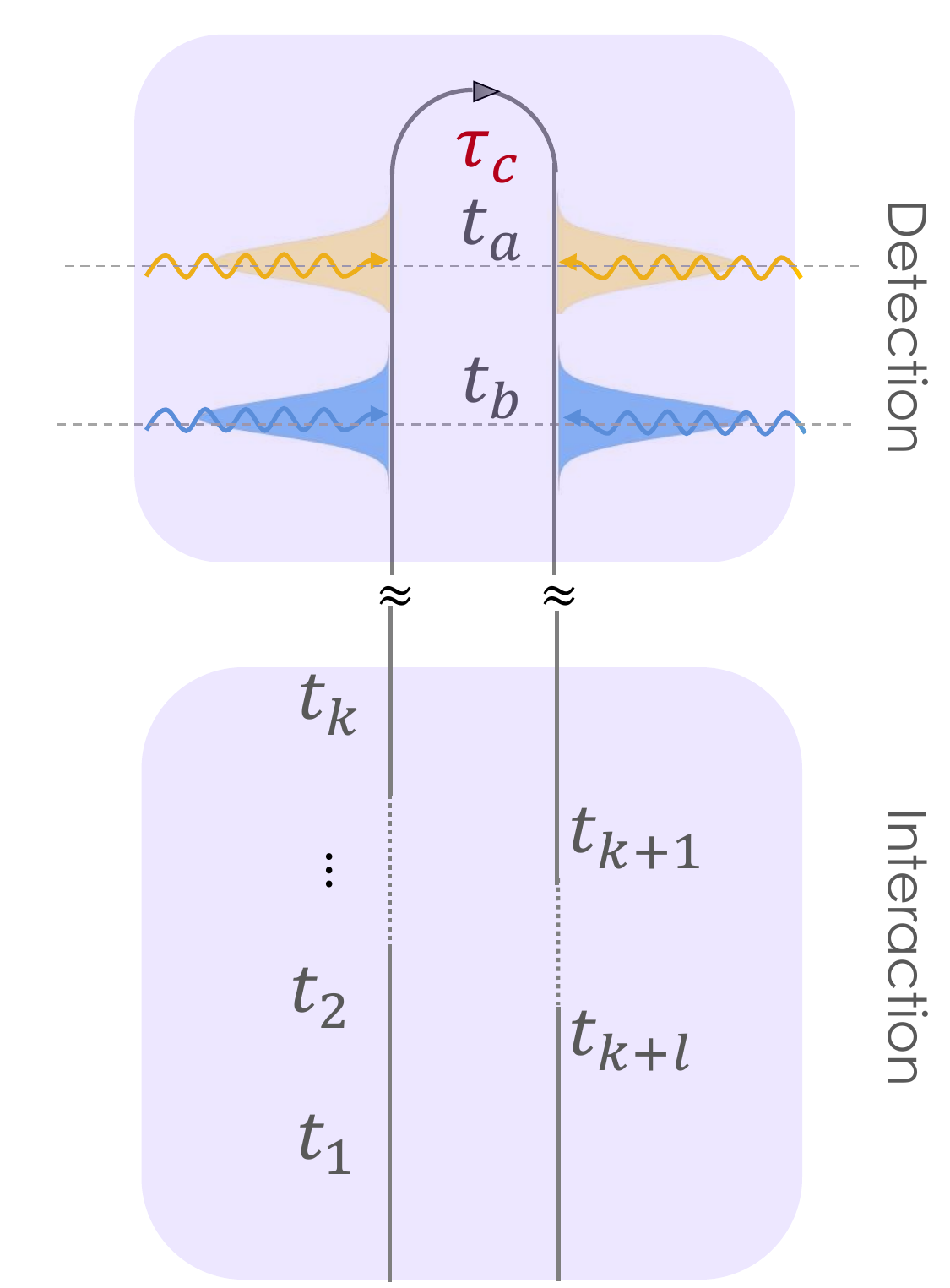}
\par\end{centering}
\caption{Diagrammatic representation of typical term in Eq. $\text{\ref{eq: Coin_Prob}}$.
The left (right) branch of the diagram represents the forward (backward)
evolution of the joint light-matter wavefunction. The number of interactions
in each branch correspond to the respective order of perturbation
of either the bra and the ket in the interaction picture. The interaction
and detection interval in this calculation do not overlap although
in principle the may. \label{fig:Diagramatic-rep}}

\end{figure}

\begin{figure}
\noindent \begin{centering}
\includegraphics[scale=0.6]{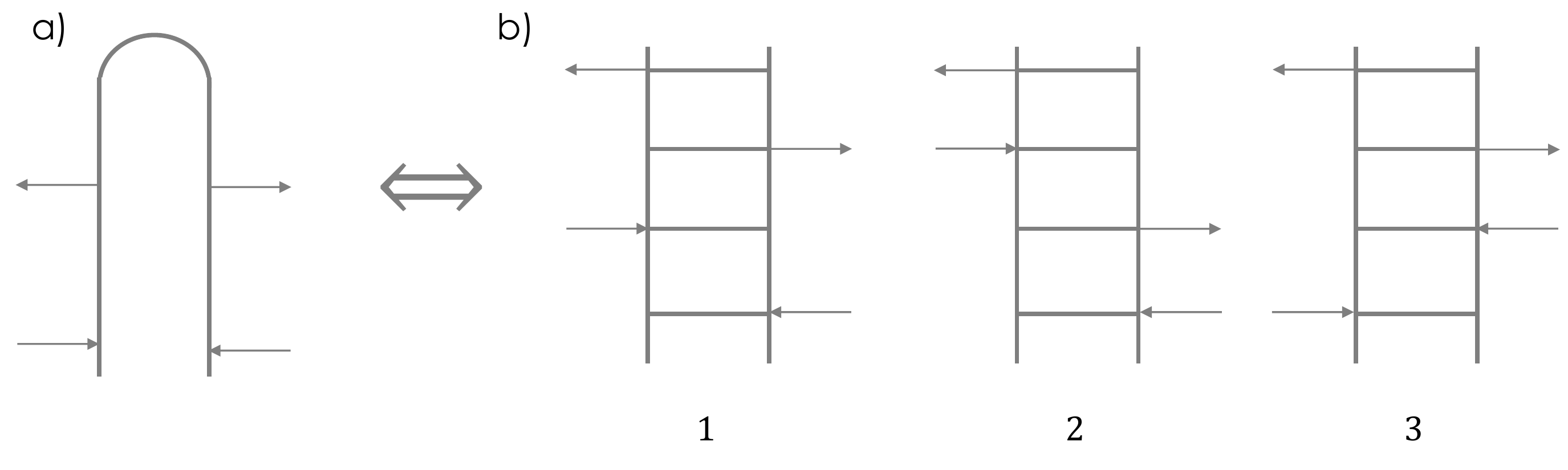}
\par\end{centering}
\caption{\textbf{Partial Vs. full time ordering}. a) a typical $4^{\text{th}}$
order term in the wavefunction pertubative treatment. b.) Similar
terms in Liouville space expansion. Breaking the loop contour (forward
and backward) into a single time evolution for both the bra and the
ket. \label{fig:Partial-Vs.-full}}

\end{figure}

Each term in Eq.$\text{\ref{eq: Coin_Prob}}$ can be represented using
a fully time ordered loop diagram as depicted in Fig. $\text{\ref{fig:Diagramatic-rep}}$.
this diagram represents the forward (in time) evolution of the ket
(left) pertubatively in the interaction picture to the $k^{\text{th}}$
order, and the bra backwards in time to the $l^{\text{th}}$ order
along the time contour $\tau_{C}$. Between interaction events, the
sample and the electromagnetic field are evolved using their free
Hamiltonian. The lowest order in which two photons interact with the
sample and detected is the $4^{\text{th}}$. Contributions with odd
number of photons from the left or right at the detection are naturally
eliminated. This corresponds to 

\begin{equation}
{\cal P}\left(\boldsymbol{r}_{a},\boldsymbol{r}_{b};t_{a},t_{b}\right)\approx\sum_{k,l=0}^{4}\langle\Psi^{\left(l\right)}\left(t\right)\vert E_{-}^{\dagger}\left(\boldsymbol{r}_{b},t_{b}\right)E_{-}^{\dagger}\left(\boldsymbol{r}_{a},t_{a}\right)E_{+}\left(\boldsymbol{r}_{a},t_{a}\right)E_{+}\left(\boldsymbol{r}_{b},t_{b}\right)\vert\Psi^{\left(k\right)}\left(t\right)\rangle_{\mu\varphi},\label{eq: approx signal}
\end{equation}

\noindent where the $\pm$ subscripts highlight the operation direction
of the field with respect to the time contour $\tau_{C}$, for positive
and negative time direction. Equation $\text{\ref{eq: approx signal}}$
gives rise to two kinds of contributions: (1) four interactions in
one side and non at the other and (2) two interactions from each side.
The Hilbert space description -- while equivalent to the alternative
Liouville space -- results in partial time ordering. The time ordering
is maintained along the contour $\tau_{C}$, thus, along the left
and right branches of the diagram in Fig. $\text{\ref{fig:Diagramatic-rep}}$individually.
Alternatively, if one is interested in evolving the density matrix
(in Liouville-space), the relative time-ordering of left and right
branches is also important, resulting in absolute time ordering. This
difference become important in the interpretation of pertubative treatments
of light-matter coupling and essential when one considers coupling
to reservoirs. As a result, in the wavefunction approach (Hilbert
space) the relative coherence during the evolution is not expressed,
only the the final phase accumulated along the entire evolution of
the bra and ket separately such that over-all coherence is accounted
for as shown in Fig. $\text{\ref{fig:Partial-Vs.-full}}$a. Alternatively,
in the Liouville-space approach the change in coherence is instantaneously
monitored in the calculation process as demonstrated in Fig.$\text{\ref{fig:Partial-Vs.-full}}$b.
For time dependent perturbation theory that includes terms which break
time reversal (bath), using the Liouville space approach is inevitable
and thus invoked in this paper. 

\section{The signal}

The signal in Eq. $\text{\ref{eq: signal}}$ is composed of all processes
$P_{i}$ evaluated with all observables $O_{j}$. One way to think
about it is that each process $P_{i}$ is obtained by coherent superposition
of all HOM detection pathways ${\cal C}_{i}\left(t,\tau,T\right)$,
and the signal is given by their superposition. The coincidence signal
in Eq. $\text{\ref{eq: signal}}$ can be written accordingly 

\begin{equation}
{\cal C}\left(t,\tau,T\right)=2\mathfrak{Re}\left\{ \sum_{i=1}^{5}{\cal S}_{\nu,i}\right\} \label{eq: signal_sup}
\end{equation}

\noindent and solved for each detection-interaction pathway combination
below separately. \\

\subsection{Example of one process-observable combination}

\begin{figure}
\begin{centering}
\includegraphics{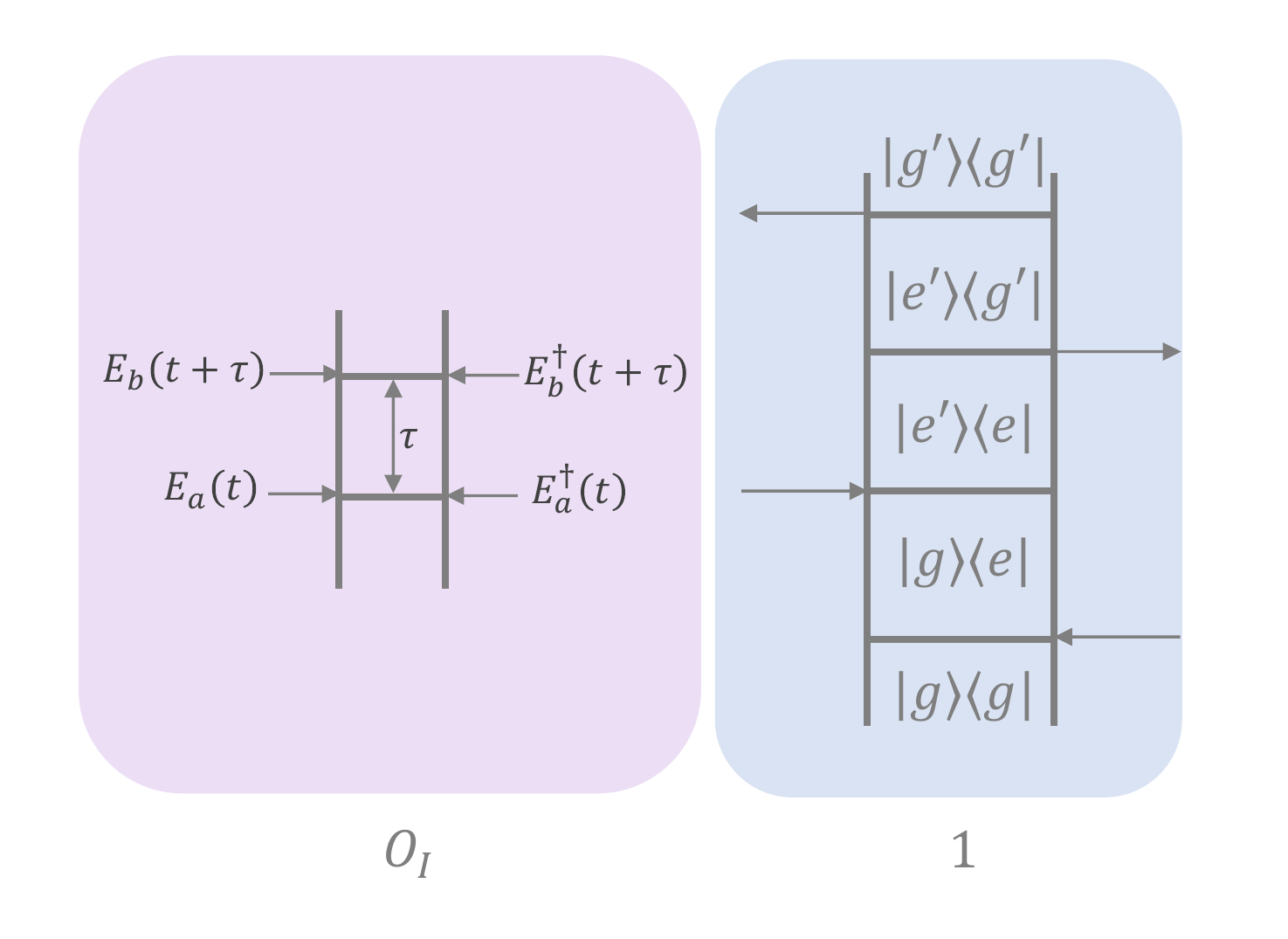}
\par\end{centering}
\caption{The resulting combination of $\nu=I$ direct detection pathway and
$i=1$ interaction pathway. \label{fig3s}}

\end{figure}

We now illustrate how to combine the preparation and observation boxes
for a single term from the total signal. We chose ${\cal S}_{I,1}$
as shown in Fig. $\text{\ref{fig3s}}$. This contribution introduces
four combinations of field modes corresponding to coupling with $a$
and $b$ modes, $\left(aa,bb,ab,ba\right)$ . We consider the realization
in which $a$ is coupled from the left and $b$ from the right. 

\begin{align*}
{\cal S}_{I,1}\vert_{a_{L}b_{R}} & =\int dt_{4}dt_{3}dt_{2}dt_{1}\theta\left(t_{1}t_{2}\right)\theta\left(t_{2}t_{3}\right)\theta\left(t_{3}t_{4}\right)\\
 & \times\text{tr}\left\{ \left[E_{b,R}^{\dagger}\left(t+\tau\right)E_{b,L}\left(t+\tau\right)E_{a,R}^{\dagger}\left(t\right)E_{a,L}\left(t\right)\right]E_{a,L}^{\dagger}\left(t_{1}\right)E_{b,R}\left(t_{2}\right)E_{a,L}\left(t_{3}\right)E_{b,R}^{\dagger}\left(t_{4}\right)\rho_{\varphi}\left(-\infty\right)\right\} \\
 & \times\text{tr}\left\{ V_{L}\left(t_{1}\right)V_{R}^{\dagger}\left(t_{2}\right)V_{L}^{\dagger}\left(t_{3}\right)V_{R}\left(t_{4}\right)\rho_{\mu}\left(-\infty\right)\right\} .
\end{align*}

The operators in the square brackets correspond to the detection process
and thus last. Initially, only two fields modes are populated and
thus we assume that after the detection process, the field returns
to its ground-state (vacuum) and obtain

\begin{align*}
{\cal S}_{I,1}\vert_{b_{L}a_{R}} & =\int dt_{4}dt_{3}dt_{2}dt_{1}\theta\left(t_{1}t_{2}\right)\theta\left(t_{2}t_{3}\right)\theta\left(t_{3}t_{4}\right)\\
 & \times\langle\text{vac}\vert E_{b,L}\left(t+\tau\right)E_{a,L}\left(t\right)E_{b,L}^{\dagger}\left(t_{1}\right)E_{b,L}\left(t_{3}\right)\vert\Psi_{\theta}\rangle_{\varphi}\langle\Psi_{\theta}\vert E_{a,R}^{\dagger}\left(t_{4}\right)E_{a,R}\left(t_{2}\right)E_{a,R}^{\dagger}\left(t\right)E_{b,R}^{\dagger}\left(t+\tau\right)\vert\text{vac}\rangle_{\varphi}\\
 & \times\left\langle V_{L}\left(t_{1}\right)V_{R}^{\dagger}\left(t_{2}\right)V_{L}^{\dagger}\left(t_{3}\right)V_{R}\left(t_{4}\right)\right\rangle _{\mu}.
\end{align*}

where we have plugged in the explicit expression for the source term
$\rho_{\varphi}\equiv\vert\Psi_{\theta}\rangle\langle\Psi_{\theta}\vert$.
using $\Phi_{\theta}\left(t_{1},t_{2}\right)=\frac{1}{\sqrt{2}}\left[\phi\left(t_{1},t_{2}\right)+e^{-i\theta}\phi\left(t_{2},t_{1}\right)\right]$.
Following Ref. \citep{Mukamel_2008}, we change the integration time
variables to time differences between interaction events and obtain

\begin{align*}
{\cal S}_{I,1}\vert_{b_{L}a_{R}} & =\int d\tau_{3}d\tau_{4}\\
 & \times\Phi_{\theta}\left(t,t-\tau_{3}\right)\Phi_{\theta}^{*}\left(t,t-\tau_{3}-\tau_{4}\right)\\
 & \times\left\langle V_{L}{\cal G}\left(\tau_{0}\right)V_{R}^{\dagger}{\cal G}\left(\tau_{2}\right)V_{L}^{\dagger}{\cal G}\left(\tau_{3}\right)V_{R}\right\rangle _{\mu},
\end{align*}

\noindent where the Liouville-space Green's function is given by ${\cal G}\left(t\right)=-\frac{i}{\hbar}\theta\left(t\right)e^{-i{\cal L}_{\mu}t-\eta t}$.
This terms was obtained from Eq. $\text{\ref{eq: signal}}$, using
one additional approximation: the free-photon propagator from the
sample to the detector is taken to be $\left\langle E_{a,L}\left(t\right)E_{a,L}^{\dagger}\left(t_{1}\right)\right\rangle \approx\delta\left(\Delta t-\nicefrac{L}{c}\right)$
where $\nicefrac{L}{c}$ is the distance between the detector and
the sample, $c$ is the speed of light and $\Delta t$ is taken to
be the time difference between the emission and detection. 

\section{The final signal -- all combinations}

Here we combine all possible contributions that correspond to all
contributing configurations of the detection and interaction pathways.
All pathways are summed as shown in Eq $\text{\ref{eq: signal}}$. 

\subsection{${\cal S}_{\nu,1}$}

\noindent 
\begin{align*}
{\cal S}_{I,1}= & \int d\tau_{3}d\tau_{4}\Phi^{*}\left(t-\tau_{3:4},t+\tau_{0}\right)\Phi\left(t,t-\tau_{3}\right){\cal F}_{1}\left(\tau_{0},\tau_{3},\tau_{4}\right)
\end{align*}

\noindent 
\begin{align*}
{\cal S}_{II,1}: & \int d\tau_{3}d\tau_{4}\Phi^{*}\left(t+\tau_{0},t-\tau_{3:4}\right)\Phi\left(t-\tau_{3},t\right){\cal F}_{1}\left(\tau_{0},\tau_{3},\tau_{4}\right)
\end{align*}

\noindent 
\begin{align*}
{\cal S}_{III,1}=- & \int d\tau_{3}d\tau_{4}\Phi^{*}\left(t,t+\tau_{0}-\tau_{3:4}\right)\Phi\left(t+\tau_{0}-\tau_{3},t-T\right){\cal F}_{1}\left(T,\tau_{3},\tau_{4}\right)\\
- & \int d\tau_{3}d\tau_{4}\Phi^{*}\left(t-\tau_{3:4},t+\tau_{0}\right)\Phi\left(t-\tau_{3},t-T\right){\cal F}_{1}\left(T+\tau_{0},\tau_{3},\tau_{4}\right)
\end{align*}

\noindent 
\begin{align*}
{\cal S}_{IV,1}=- & \int d\tau_{3}d\tau_{4}\Phi^{*}\left(t+T+\tau,t-T-\tau_{3:4}\right)\Phi\left(t-T-\tau_{3},t+\tau_{0}\right){\cal F}_{1}\left(T,\tau_{3},\tau_{4}\right)\\
- & \int d\tau_{3}d\tau_{4}\Phi^{*}\left(t+T+\tau,t-T-\tau_{3:4}\right)\Phi\left(t,t-T-\tau_{3}\right){\cal F}_{1}\left(\tau_{0}+T,\tau_{3},\tau_{4}\right)
\end{align*}

\subsection{${\cal S}_{\nu,2}$}

\noindent 
\begin{align*}
{\cal S}_{I,2}= & \int d\tau_{3}d\tau_{4}\Phi^{*}\left(t-\tau_{4},t+\tau_{0}\right)\Phi\left(t,t+\tau_{3}\right){\cal F}_{2}\left(\tau_{0}-\tau_{3},\tau_{3},\tau_{4}\right)
\end{align*}

\noindent 
\begin{align*}
{\cal S}_{II,2}= & \int d\tau_{3}d\tau_{4}\Phi^{*}\left(t+\tau_{0},t-\tau_{4}\right)\Phi\left(t+\tau_{3},t\right){\cal F}_{2}\left(\tau_{0}-\tau_{3},\tau_{3},\tau_{4}\right)
\end{align*}

\noindent 
\begin{align*}
{\cal S}_{III,2} & =-\int d\tau_{3}d\tau_{4}\Phi^{*}\left(t,t+\tau_{0}-\tau_{4}\right)\Phi\left(t+\tau_{0}+\tau_{3},t-T\right){\cal F}_{2}\left(T-\tau_{3},\tau_{3},\tau_{4}\right)\\
 & -\int d\tau_{3}d\tau_{4}\Phi^{*}\left(t-\tau_{4},t+\tau_{0}\right)\Phi\left(t+\tau_{3},t-T\right){\cal F}_{2}\left(\tau_{0}+T-\tau_{3},\tau_{3},\tau_{4}\right)
\end{align*}

\noindent 
\begin{align*}
{\cal S}_{IV,2}=- & \int d\tau_{3}d\tau_{4}\Phi^{*}\left(t+T+\tau,t-T-\tau_{4}\right)\Phi\left(t-T+\tau_{3},t+\tau_{0}\right){\cal F}_{2}\left(T-\tau_{3},\tau_{3},\tau_{4}\right)\\
- & \int d\tau_{3}d\tau_{4}\Phi^{*}\left(t+T+\tau,t-T-\tau_{4}\right)\Phi\left(t,t-T+\tau_{3}\right){\cal F}_{2}\left(\tau_{0}+T-\tau_{3},\tau_{3},\tau_{4}\right)
\end{align*}

\subsection{${\cal S}_{\nu,3}$}

\noindent 
\begin{align*}
{\cal S}_{I,3}= & \int d\tau_{3}d\tau_{4}\Phi^{*}\left(t-\tau_{3},t+\tau_{0}\right)\Phi\left(t,t-\tau_{3:4}\right){\cal F}_{3}\left(\tau_{0},\tau_{3},\tau_{4}\right)
\end{align*}

\noindent 
\begin{align*}
{\cal S}_{II,3}= & \int d\tau_{3}d\tau_{4}\Phi^{*}\left(t+\tau_{0},t-\tau_{3}\right)\Phi\left(t-\tau_{3:4},t\right){\cal F}_{3}\left(\tau_{0},\tau_{3},\tau_{4}\right)
\end{align*}

\noindent 
\begin{align*}
{\cal S}_{III,3}=- & \int d\tau_{3}d\tau_{4}\Phi^{*}\left(t,t+\tau_{0}-\tau_{3}\right)\Phi\left(t+\tau_{0}-\tau_{3:4},t-T\right){\cal F}_{3}\left(T,\tau_{3},\tau_{4}\right)\\
- & \int d\tau_{3}d\tau_{4}\Phi^{*}\left(t-\tau_{3},t+\tau_{0}\right)\Phi\left(t-\tau_{3:4},t-T\right){\cal F}_{3}\left(\tau_{0}+T,\tau_{3},\tau_{4}\right)
\end{align*}

\noindent 
\begin{align*}
{\cal S}_{IV,3}=- & \int d\tau_{3}d\tau_{4}\Phi^{*}\left(t+T+\tau,t-T-\tau_{3}\right)\Phi\left(t-T-\tau_{3:4},t+\tau_{0}\right){\cal F}_{3}\left(T,\tau_{3},\tau_{4}\right)\\
- & \int d\tau_{3}d\tau_{4}\Phi^{*}\left(t+T+\tau,t-T-\tau_{3}\right)\Phi\left(t,t-T-\tau_{3:4}\right){\cal F}_{3}\left(T+\tau_{0},\tau_{3},\tau_{4}\right)
\end{align*}

\subsection{${\cal S}_{\nu,4}$}

\noindent 
\begin{align*}
{\cal S}_{I,4}= & \int d\tau_{3}d\tau_{4}\Phi^{*}\left(t,t+\tau_{0}\right)\Phi\left(t-\tau_{4},t+\tau_{3}\right){\cal F}_{4}\left(\tau_{0}-\tau_{3},\tau_{3},\tau_{4}\right)
\end{align*}

\noindent 
\begin{align*}
{\cal S}_{II,4} & =\int d\tau_{3}d\tau_{4}\Phi^{*}\left(t+\tau_{0},t\right)\Phi\left(t+\tau_{0}+\tau_{3},t+\tau_{0}-\tau_{4}\right){\cal F}_{4}\left(\tau_{0}-\tau_{3},\tau_{3},\tau_{4}\right)
\end{align*}

\noindent 
\begin{align*}
{\cal S}_{III,4}=- & \int d\tau_{3}d\tau_{4}\Phi^{*}\left(t,t+\tau_{0}\right)\Phi\left(t-T+\tau_{3},t-T-\tau_{4}\right){\cal F}_{4}\left(2T+\tau_{0}-\tau_{3},\tau_{3},\tau_{4}\right)
\end{align*}

\noindent 
\begin{align*}
{\cal S}_{IV,4}=- & \int d\tau_{3}d\tau_{4}\Phi^{*}\left(t+T+\tau,t-T\right)\Phi\left(t-\tau_{4},t+\tau_{3}\right){\cal F}_{4}\left(\tau_{0}-\tau_{3},\tau_{3},\tau_{4}\right)
\end{align*}

\subsection{${\cal S}_{\nu,5}$}

Here, each contribution is naturally symmetrized. 

\noindent 
\[
{\cal S}_{I,5}=\int d\tau_{3}d\tau_{4}\Phi^{*}\left(t,t+\tau_{0}\right)\left[\Phi_{ab}\left(t-\tau_{3},t-\tau_{3:4}\right)+\Phi_{ba}\right]{\cal F}_{5}\left(\tau_{0},\tau_{3},\tau_{4}\right)
\]

\noindent 
\[
{\cal S}_{II,5}=\int d\tau_{3}d\tau_{4}\Phi^{*}\left(t+\tau_{0},t\right)\left[\Phi_{ab}\left(t-\tau_{3},t-\tau_{3:4}\right)+\Phi_{ba}\right]{\cal F}_{5}\left(\tau_{0},\tau_{3},\tau_{4}\right)
\]

\noindent 
\[
{\cal S}_{III,5}=-\int d\tau_{3}d\tau_{4}\Phi^{*}\left(t,t+\tau_{0}\right)\left[\Phi_{ab}\left(t-T-\tau_{3:4},t-T-\tau_{3}\right)+\Phi_{ba}\right]{\cal F}_{5}\left(\tau_{0}+2T,\tau_{3},\tau_{4}\right)
\]

\noindent 
\[
{\cal S}_{IV,5}=-\int d\tau_{3}d\tau_{4}\Phi^{*}\left(t+T+\tau_{0},t-T\right)\left[\Phi_{ab}\left(t-\tau_{3},t-\tau_{3:4}\right)+\Phi_{ba}\right]{\cal F}_{5}\left(\tau_{0},\tau_{3},\tau_{4}\right)
\]

\end{document}